\documentclass[10pt,preprint]{aastex}

\newcommand{\mj}{$M_{\rm Jup}$}
\newcommand{\msun}{$M_{\odot}$}

\begin{document}

\shorttitle{Search for Companions to Stars with Planets}
\shortauthors{Luhman \& Jayawardhana}

\title{An Adaptive Optics Search for Companions to Stars with 
Planets}


\author{K. L. Luhman}
\affil{Harvard-Smithsonian Center for Astrophysics, 60 Garden St., 
Cambridge, MA 02138}

\email{kluhman@cfa.harvard.edu}

\and

\author{Ray Jayawardhana}
\affil{Department of Astronomy, University of California, 601 Campbell Hall,
Berkeley, CA 94720}

\email{rayjay@astro.berkeley.edu}

\begin{abstract}
We have performed a Keck adaptive optics (AO) imaging survey of 25
extrasolar planetary systems discovered by the radial velocity programs.
Typically, the high-resolution (FWHM$\sim0\farcs04$) near-infrared images 
are able to detect point sources at $\Delta H=10$ at 
$1\arcsec$, L and T dwarfs at $0\farcs5$ to $2\arcsec$, and stars and brown 
dwarfs at 3-10~AU and 10-100~AU from the targets.
The AO images reveal 15 faint sources ($H=14$-20) near the stars HD~37124, 
HD~168443, HD~187123, and HD~190228 out to a search radius of $3\farcs3$. 
We have obtained AO images at a second epoch for five of the candidate 
companions at Keck and Gemini Observatories. The resulting measurements of 
proper motions relative to the primary stars indicate that these five sources 
are background stars. 
Given that the four primaries are near the Galactic plane ($|b|<6\arcdeg$), 
the remaining faint sources are also likely to be background stars rather than 
substellar companions. We present the companion detection limits for 
each target as a function of separation and compare them to the
sensitivities of radial velocity surveys, demonstrating the 
complementary nature of the two techniques. 

\end{abstract}


\section{Introduction}

In the search for planetary companions outside of our solar system, a variety
of techniques have been devised and implemented, of which radial velocity 
monitoring has proven the most successful to date.
Such measurements have led to the discovery of planets with masses 
of $>0.3$~\mj\ and separations of a few AU or less around $\sim60$ nearby stars 
(Mayor \& Queloz 1995; Marcy \& Butler 1996; Cochran et al.\ 1997; Noyes et 
al.\ 1997; Udry, Mayor, \& Queloz 2001; Butler et al.\ 2000; 
Fischer et al.\ 2001b; Tinney et al.\ 2001).
Combining those results with data on spectroscopic binaries produces a
distribution of companions from 1-1000~\mj\ that is characterized by 
a minimum or ``brown dwarf desert" at 10-30~\mj\ (e.g., Marcy \& Butler 1998).
This feature is likely a reflection of two formation mechanisms, one that 
creates planets from disks and one that produces companion stars and brown 
dwarfs (e.g., core fragmentation). However, to fully understand the processes
of planet and binary star formation, it is essential that the 
frequency of companions is measured over the maximum possible range of 
separations. For instance, the conspicuous paucity of brown dwarf companions at 
$\lesssim5$~AU may not exist at separations beyond 1000~AU (Gizis et al.\ 2001).

Known extrasolar planetary systems are logical initial targets
for a search for companions at modestly wide separations (10-100~AU).
Gravitational perturbations by outer substellar or stellar companions 
may be responsible for the eccentric orbits observed for many planetary
companions. To properly evaluate the
stability and evolution of these planets (e.g., Rivera \& Lissauer 2000),
wide massive companions must be searched for and included in the analysis.
Furthermore, if substellar companions can be resolved from primaries,
spectroscopic analysis would provide constraints on masses, temperatures, 
surface gravities, and compositions, and thus contribute to our understanding 
of the physical characteristics of substellar objects (Marley et al.\ 1996;
Allard et al.\ 1996).
Finally, as radial velocity measurements are accumulated over longer periods 
of time, they are becoming sensitive to companions at larger separations. 
It already appears 
that at least 50\% of stars with planets have additional distant companions
(Marcy et al.\ 2001a, b; Fischer et al.\ 2001b), which suggests that 
a search for a substellar objects at still wider separations through
high-resolution direct imaging could prove fruitful.

Direct imaging has become increasingly effective in detecting faint
companions to bright stars.  In high-resolution imaging with
WFPC2 aboard the {\it Hubble Space Telescope (HST)}, Schroeder et al.\ (2000)
probed for massive and young brown dwarfs at 1-60~AU from 23 nearby stars.  
From the ground, Oppenheimer et al.\ (2001) used optical coronographic data and 
near-infrared (IR) images to search for companions at 40-120~AU and
$>40$~\mj\ near most of the known stars within 8~pc, which resulted in the 
discovery of the brown dwarf companion Gl~229B (Nakajima et al.\ 1995).
The tip-tilt techniques used for enhanced image quality by 
Oppenheimer et al.\ (2001) have been followed by higher order adaptive optics
(AO) systems that achieve comparable and sometimes superior spatial resolution
and sensitivity to those of $HST$. Near-IR AO imaging already has been
used to discover binary systems whose members are near the hydrogen
burning mass limit (Mart{\'\i}n et al.\ 2000; Close et al.\ 2002). 
In addition, AO has been performed in conjunction with radial velocity
measurements of nearby stars (Delfosse et al.\ 1999) and has resulted in 
the discovery of a brown dwarf companion ($\sim50$~\mj, $H=14.4$) to the
planet-bearing star Gl~86 (Els et al.\ 2001).
In this paper, we present the results of a survey for companions to 25
planetary systems using near-IR AO imaging at Keck Observatory. We report
astrometry and photometry for several point sources found in these images
and discuss their likely origin as companions or background field stars.
We characterize the detection limits of these data in both observational
and physical units, which are then compared to the sensitivities of the
radial velocity surveys, and we discuss the implications of these data 
for individual planet-bearing stars.
 
\section{Observations and Data Analysis}
\label{sec:obs}

\subsection{Adaptive Optics Imaging}

During the nights of 2000 May 10, June 27, and September 16-18, we 
used the Shack-Hartmann AO system on the Keck~II telescope 
(Wizinowich et al.\ 2000) in conjunction with KCam, a NICMOS3 $256\times256$
near-IR camera, to obtain images of 25 stars for which planetary companions
have been discovered in previous radial velocity measurements. Properties
of the planetary systems in our sample are given in Table~\ref{tab:prop}.
The plate scale of KCam was $0\farcs01744\pm0\farcs0005$~pixel$^{-1}$, 
corresponding to a total field of $4\farcs46\times4\farcs46$. The lower right
quadrant of the array was not functional.
A cold filter wheel contained $J$, $H$, and $K\arcmin$ filters. 
Warm external filters included an open filter and neutral density filters that 
attenuated the signal by factors of 10, 100, and 1000 (N1, N2, and N3).
The H-band filter was selected for all observations 
because it produces the optimum combination of spatial resolution and 
sensitivity and because substellar objects are relatively bright in this band. 

The targets were grouped into pairs of stars that were close together on the
sky ($\lesssim15\arcdeg$). The stars in a pair were observed in succession 
so that the data for one star provided an estimate of the PSF for the other 
star. Images for a given star were obtained in the following manner. 
The star was first centered in one of the three operating quadrants of the 
array. The exposure time and neutral density filter were selected such that
the maximum number of counts in the image was between 50 and 80\% of the 
saturation level, where 5~sec and N3 were typical choices for a star at $H=5$. 
Multiple exposures were obtained at each of two positions that were separated 
by $0\farcs0872$ (5 pixels) along the rows and columns of the array. Similar 
exposures were obtained through the less opaque neutral density filters 
and the open filter to reach successively fainter magnitudes.
The same steps were then followed for the star that was paired to this target.
These data are used to search for companions within the quadrant surrounding
each star ($<1\farcs1$). The following observations were designed to reach 
larger separations. A target was centered in the upper left quadrant of the
camera array.
Because the area of the array directly surrounding the target was 
already imaged and would be saturated in long exposures,
the telescope was offset by $2\farcs232$ (128 pixels) in each 
direction on the array to place the star near the center of the 
lower right dead quadrant. Three images were obtained at each position
in the same dither pattern used previously. 
Each image consisted of the sum of two exposures with integration times 
of 30~sec for most stars and 5~sec for the brightest targets, which were 
short enough to avoid saturation within the operating quadrants of the
array. As a result, the total integration time was typically 6~min.
These steps for the long exposures were repeated 
at two additional position angles at intervals of $120\arcdeg$, providing
full and partial coverage out to radii of $3\farcs3$ and $4\farcs6$ 
surrounding each star. The position angle of the array was fixed during
a given image.

We now comment on a few departures from the above observing strategy.
The observations of $\rho^1$~55~Cnc and 47~UMa differed from the standard
procedure because the observations of these two objects were completed before 
the final strategy was adopted. Only the N3 and open external filters were used
for these stars. In addition, only short exposure times were used for 47~UMa.
Similarly, only short exposures were obtained for HD~168443 on the first night
of observations. As a result, only the brightest two objects near HD~168443
in Table~\ref{tab:cand} were detected in those data. A complete data set 
with the final observing procedure was obtained for HD~168443
during the later observing runs. 
The faint objects near the HD187123, HD37124, and HD168443 were noticed 
in the data that were reduced at the telescope. To
facilitate the measurement of the positions of these sources relative to the
primaries, we obtained additional exposures in which both the faint sources
and the primaries appeared in good quadrants of the array.

Two of the stars with faint nearby sources in the Keck data, HD~187123 and
HD~37124, were observed with the University of Hawaii Hokupa'a AO system 
in conjunction with the $1024\times1024$ near-IR camera QUIRC (Graves et 
al.\ 1998) on the Gemini North telescope during the night of 2001 October 7. 
The plate scale of QUIRC was $0\farcs01998\pm0\farcs0008$~pixel$^{-1}$,
corresponding to a total field of $20\farcs46\times20\farcs46$.
We obtained twelve 30~sec exposures of HD~187123 and seven 1~sec exposures 
of HD~37124 through the $H$-band filter. These exposure times were selected
to minimize the saturation of the primary stars while providing sufficient
sensitivity to detect the faint sources found in the Keck data, 
thus optimizing the astrometry for those sources relative to the primaries.
The Gemini AO data for HD~187123 and HD~37124 exhibited FWHM$=0\farcs14$ and
$0\farcs09$, respectively.

\subsection{Image Processing}

Standard data reduction procedures were followed for the AO images.
The Keck images were divided by twilight sky flat frames, while the Gemini
data were divided by dome flat frames.
Offsets between dithered frames were measured from the centroids of 
nonsaturated point sources or the isophotes around saturated stars. 
Dithered frames were then combined into one image.

\begin{figure}
\plotone{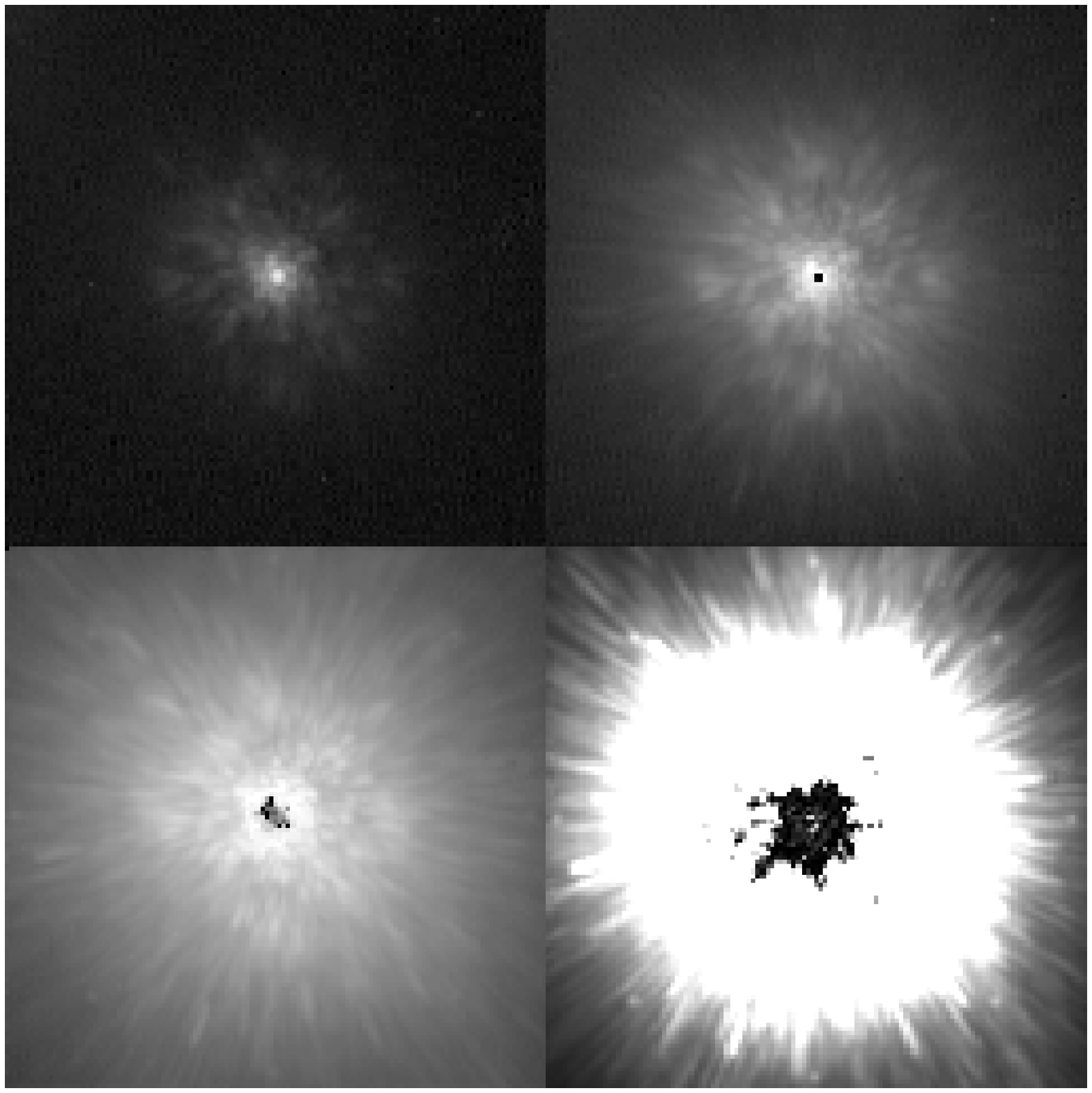}
\caption{
Keck AO images of 51~Peg ($2\farcs11\times2\farcs11$) through the $H$-band
filter and the N3, N2, N1, and open filters (upper left through lower right).
The displays of the first three images are scaled logarithmically from zero 
counts to the saturation level. In the image with the open filter,
eight artificial stars have been inserted at intervals of $45\arcdeg$ 
surrounding the primary to illustrate the point source detection limit of 
$\Delta H=10$ at $1\arcsec$. The display is scaled linearly across a range 
that is optimized for viewing the artificial stars.
Each image is the average of six 2~sec exposures. 
}
\label{fig:mosaic}
\end{figure}

\begin{figure}
\plotone{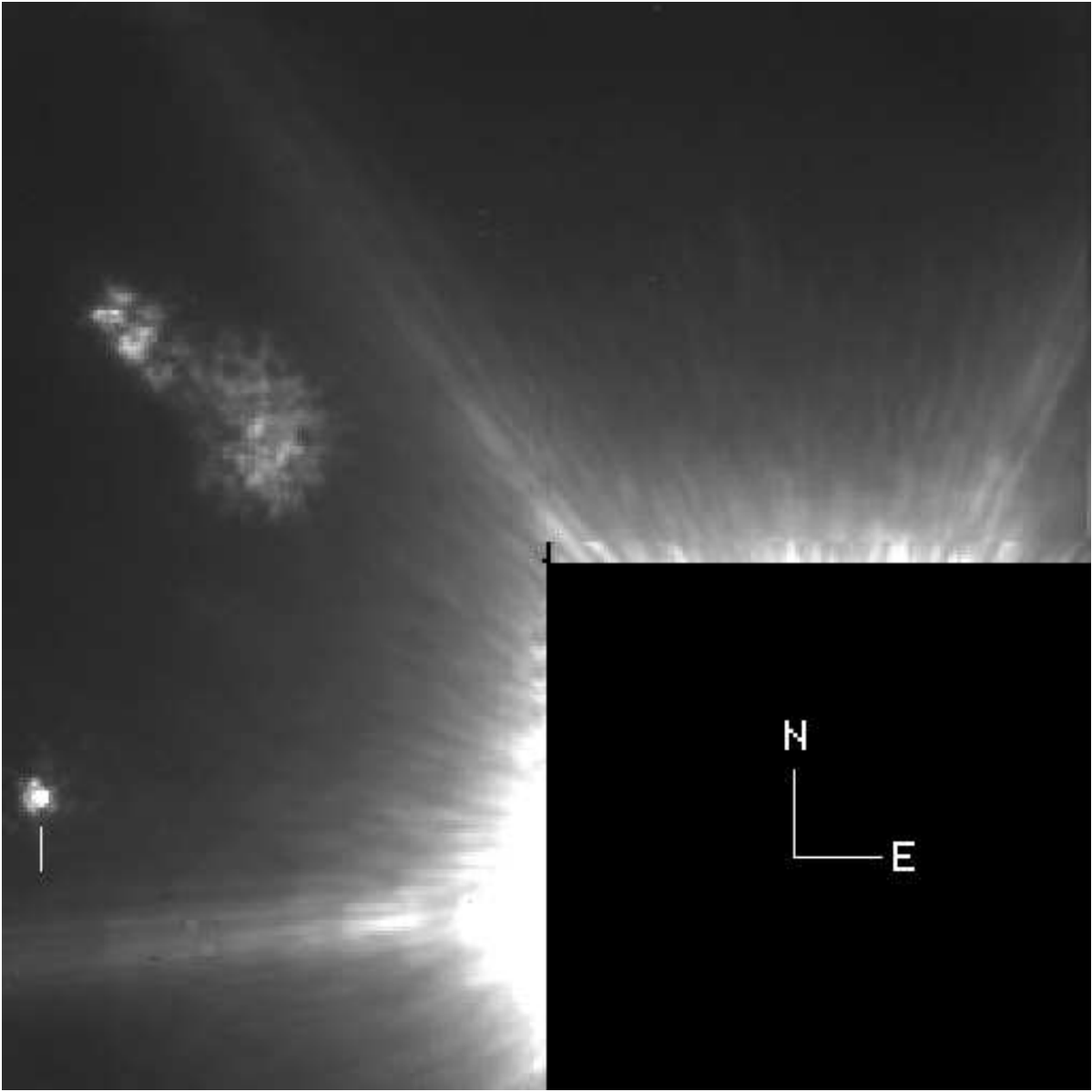}
\caption{
A Keck AO image towards HD~37124 ($4\farcs36\times4\farcs36$) through the
$H$-band filter.
A point source is detected in the lower left quadrant of the image ($H=14.9$).
The proper motion of this source relative to HD~37124 is indicative of
a background star. 
The image is the average of six 30~sec exposures and is displayed with 
a linear scale. A ghost image from HD~37124 appears in the upper left quadrant.
}
\label{fig:hd37124}
\end{figure}

\begin{figure}
\plotone{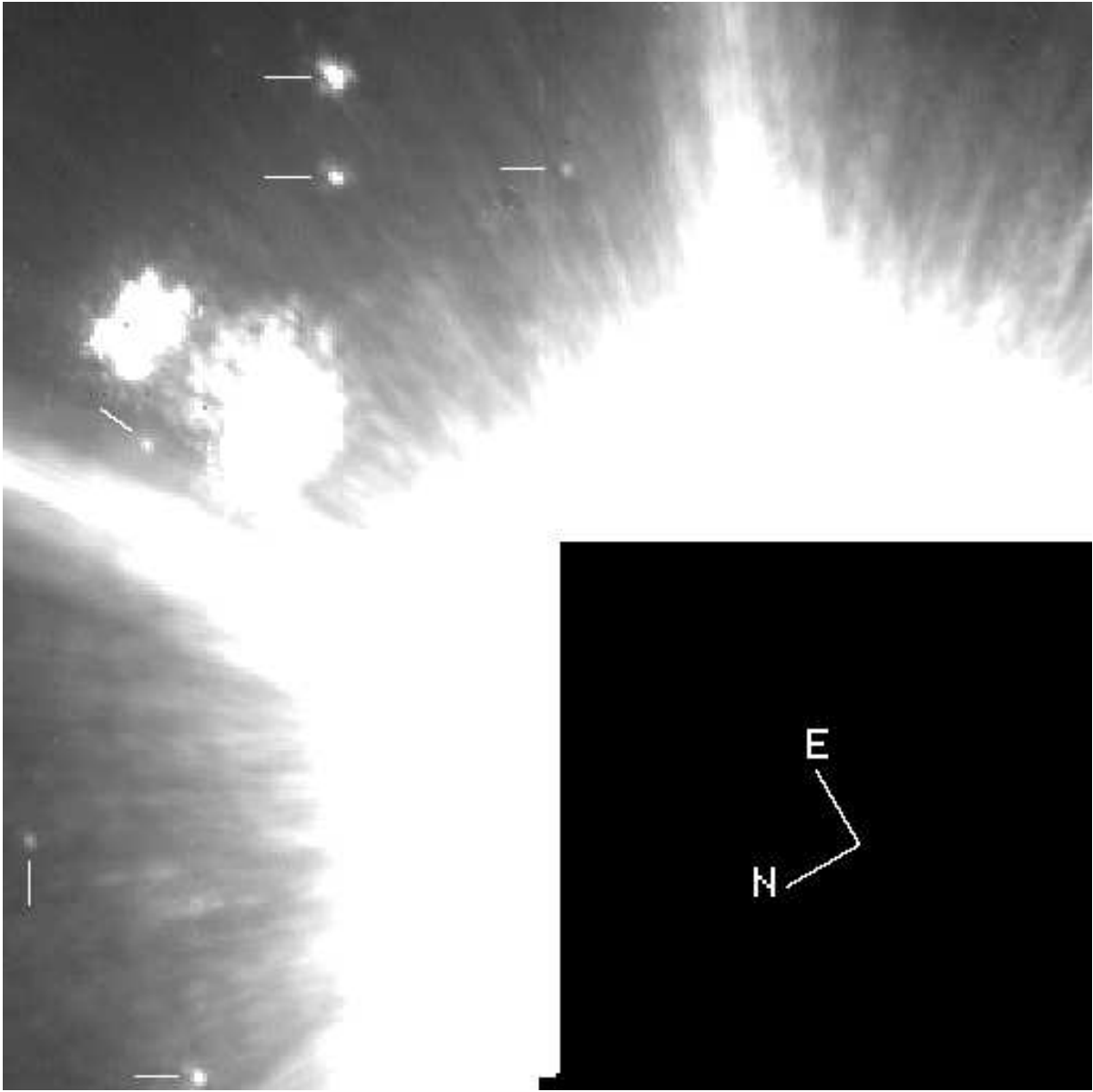}
\caption{
A Keck AO image towards HD~168443 ($4\farcs36\times4\farcs36$) through the
$H$-band filter.
Six point sources are detected in this image ($H=16.4$-19).
The ghost image from HD~168443 in the upper left quadrant falls on top of 
a seventh point source that is detected in other images of this star.
Proper motions relative to HD~168443 are available for only the two 
brightest objects; those measurements are indicative of background stars. 
The image is the average of six 30~sec exposures. The display is scaled 
linearly across a range that is optimized for viewing the six point sources. 
}
\label{fig:hd168443a}
\end{figure}

\begin{figure}
\plotone{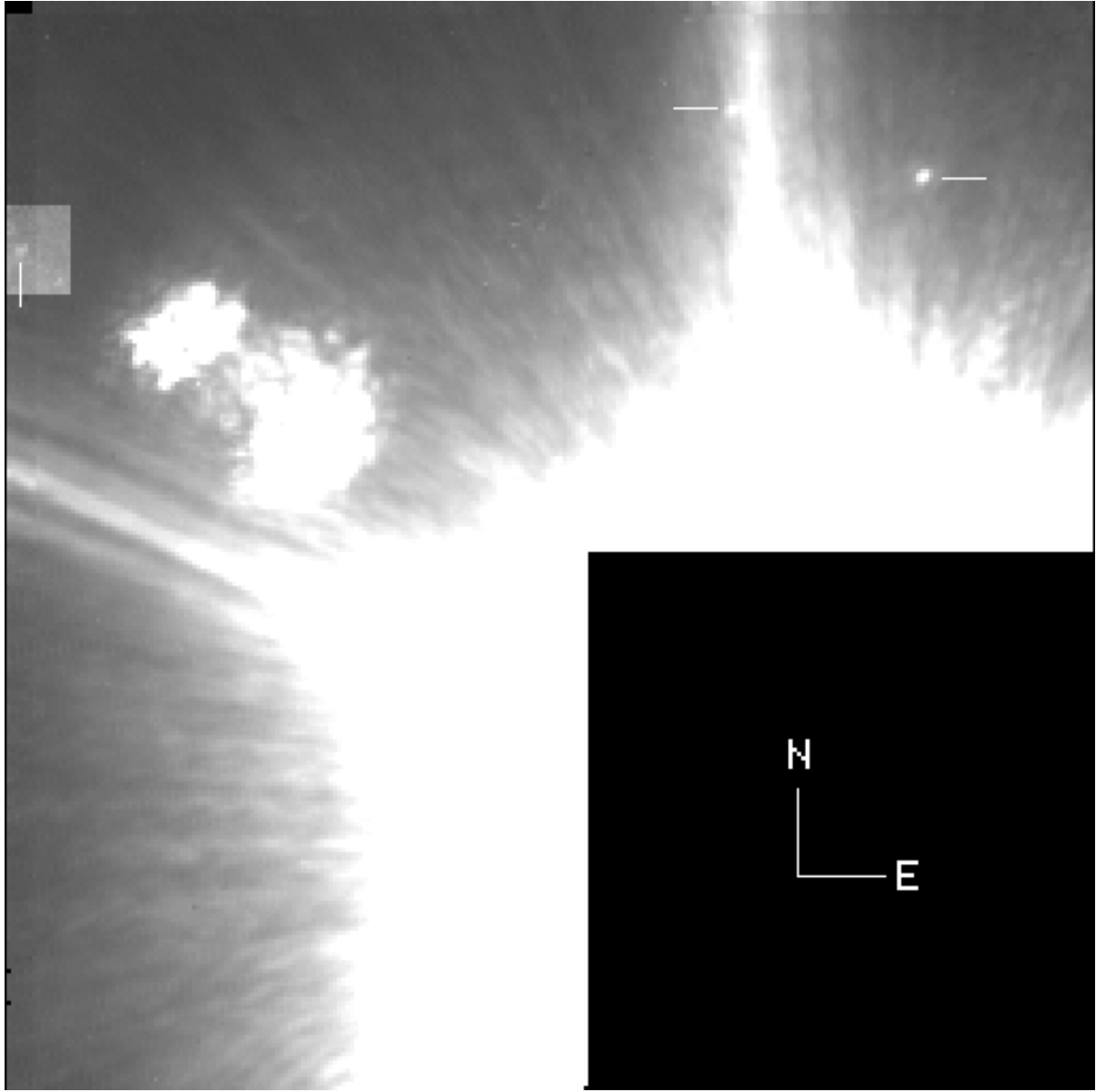}
\caption{
A Keck AO image towards HD~168443 ($4\farcs36\times4\farcs36$) through the
$H$-band filter.
Three point sources are detected in the image ($H=17.6$-19).
The image is the average of six 30~sec exposures. The display is scaled 
linearly across a range that is optimized for viewing the three point sources,
where a separate scaling is used for the area surrounding the faintest object
at the far left.
A ghost image from HD~168443 appears in the upper left quadrant.
}
\label{fig:hd168443b}
\end{figure}

\begin{figure}
\plotone{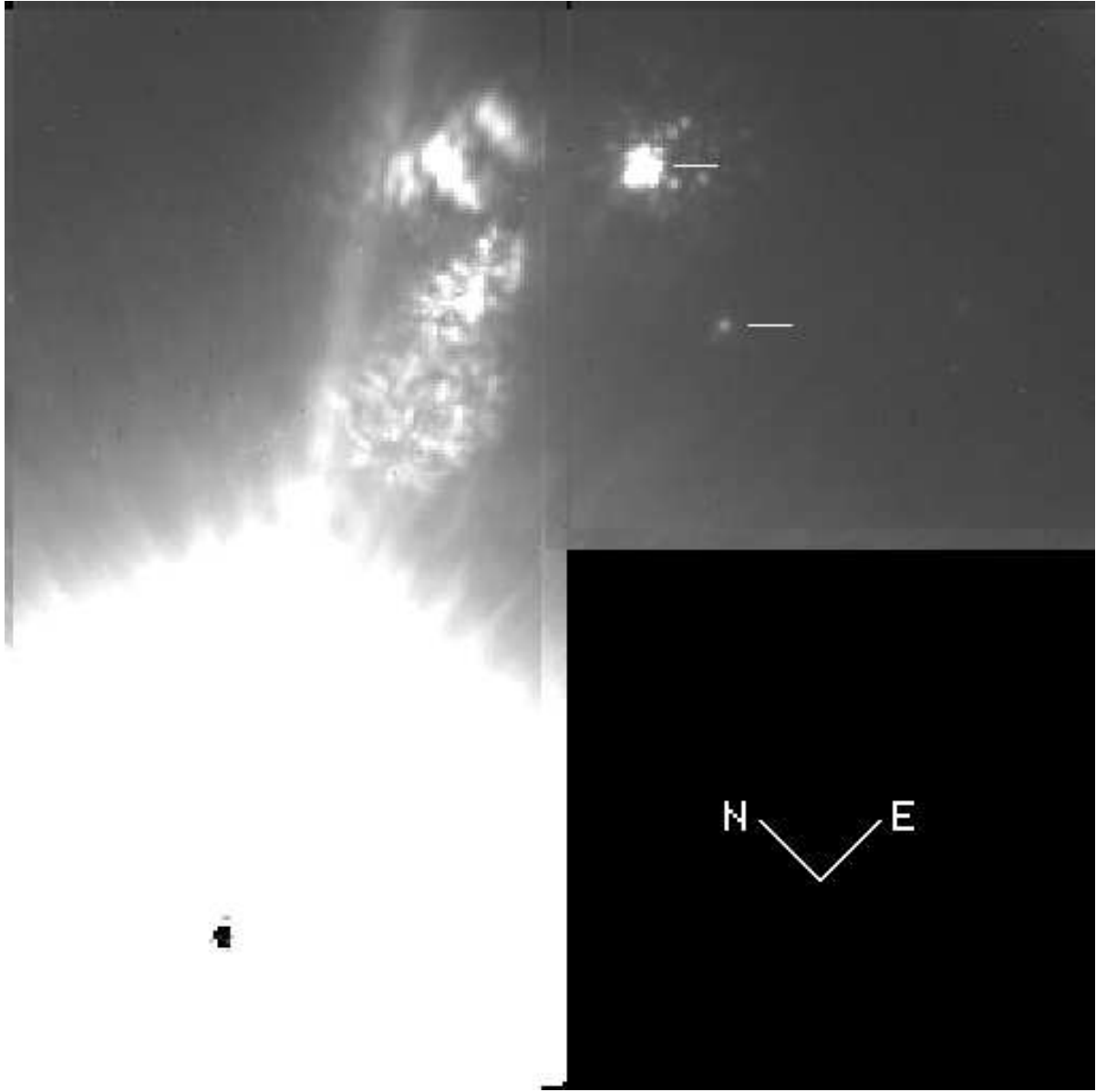}
\caption{
A Keck AO image towards HD~187123 ($4\farcs36\times4\farcs36$).
Two point sources are detected in the upper right quadrant ($H=14.2$ and 18.3). 
The proper motions
of these sources relative to HD~187123 are indicative of background stars. 
The image is the average of six 30~sec exposures. The display is scaled 
linearly across a range that is optimized for viewing the two point sources. 
A ghost image from HD~187123 appears in the upper left quadrant.
}
\label{fig:hd187123}
\end{figure}

\begin{figure}
\plotone{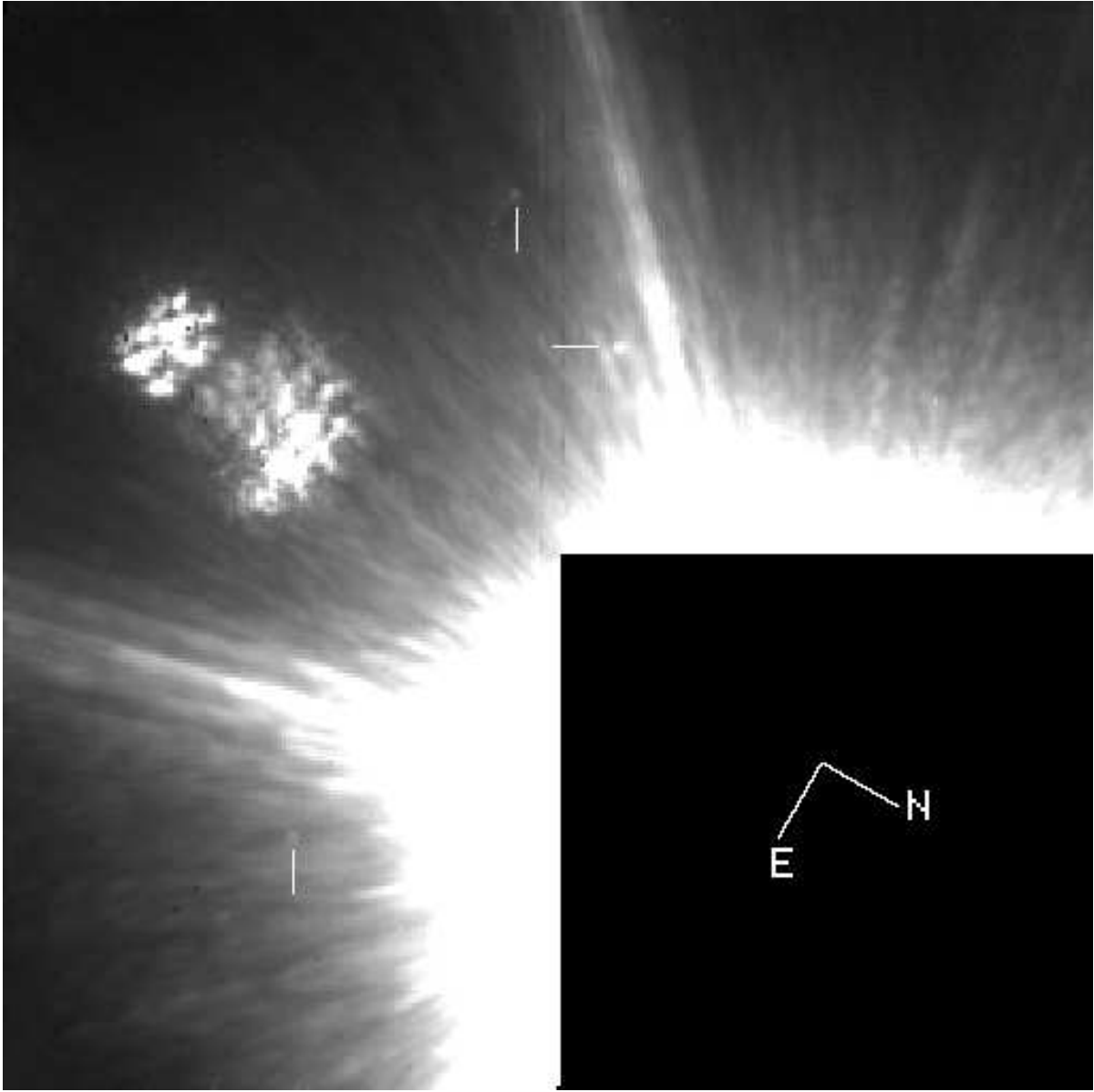}
\caption{
A Keck AO image towards HD~190228 ($4\farcs36\times4\farcs36$).
Three point sources are detected in the image ($H=18.5$-19).
The image is the average of six 30~sec exposures.  The display is scaled 
linearly across a range that is optimized for viewing the three point sources. 
A ghost image from HD~190228 appears in the upper left quadrant.
}
\label{fig:hd190228}
\end{figure}

The properties of the Keck AO data are illustrated in Figure~\ref{fig:mosaic},
which shows images of 51~Peg at various signal levels. The AO-corrected 
PSF is characterized by a bright narrow core whose width 
(FWHM$\sim0\farcs04=2.3$~pixels)
approaches the diffraction limit and a low-level broad halo that is comparable 
in size to the seeing disc (FWHM$\sim0\farcs5=29$~pixels). 
Astronomical point sources must be distinguished from point-like speckles that
reside within the halo. This was accomplished by visually comparing the 
images of a given pair of targets that were observed consecutively and thus 
act as PSF stars for each other. In addition, because most of the PSF features
rotate on the array (and thus the sky) with time, we could distinguish
speckles from astronomical sources by comparing the
images at the individual dither positions for a star. At larger radii from
a star, the features of the PSF are radially elongated and therefore 
are not mistaken for companions. The sources that were found in this
manner are listed in Table~\ref{tab:cand}.
Images of the objects detected near HD~37124, HD~168443, HD~187123, and 
HD~190228 are shown in in Figs.~\ref{fig:hd37124}-\ref{fig:hd190228}.

The array coordinates and photometry of the point sources near the target
stars were measured with the IRAF tasks IMEXAMINE and PHOT, respectively.
Aperture photometry was extracted with a radius of 3 to 6 pixels in the Keck
data, where fainter sources were measured with smaller radii. 
Photometric calibration was derived from aperture photometry with radii of
40 pixels of the unsaturated images of the target stars. When available, 
the $H$-band measurements from the Two-Micron All-Sky Survey were
adopted. Otherwise, the $H$ magnitudes were estimated by combining the 
spectral types and visual magnitudes of the targets. Most of the Keck 
images were obtained in photometric conditions and the calibrations implied 
by the various targets agree within $\pm0.1$~mag. Because the conditions
were not photometric during the Gemini observations, photometry is not measured
from those data. After measuring the aperture corrections between 40 and 
3-6 pixels from the unsaturated Keck images of the targets, we
arrived at $H$-band magnitudes for the point sources, which are given in
Table~\ref{tab:cand}. Because the fraction of the star's light that falls 
within the PSF core can change significantly between consecutive images, 
the absolute photometry has large uncertainties of $\sim\pm0.5$~mag. 
Multiple observations of the point source near HD~37124 suggest a smaller
photometric error of $\sim\pm0.2$~mag. Because the two sources near HD~187123
were measured in the same image, their relative photometry should have good
precision ($\sim\pm0.1$~mag). Similarly, the relative positions of multiple 
sources near a star have small errors ($\pm0.25$~pixels). Meanwhile,
the offsets of the faint sources from the central stars are more uncertain
because the latter were saturated or in the dead quadrant in the images 
where the former were detected. As described in the previous section, extra 
images were obtained for HD~37124, HD~187123, and HD~168443 so that the faint
sources and the primaries would both appear in good quadrants of the array. 
The uncertainties in the offsets between saturated primaries and the 
surrounding objects are $\pm1$~pixel in each direction on the array. 
In the Gemini data, both the 
candidate companions and the primaries fell within the array as well.
For the faint sources near HD~190228, we have only the original long exposures
in which the primary fell in the dead quadrant. In these data, we used the 
isophotes of the primary's PSF that extended into the good quadrants to 
estimate the position of the primary, and thus the offset between the faint 
sources and the primary. The uncertainties in those offsets are $\pm2$~pixels.
By obtaining images of an artificial light source within the Keck AO system at
several positions across the array, we measured a plate scale of 
$0\farcs01744$~pixel$^{-1}$ for KCam. A plate scale of 
$0\farcs01998$~pixel$^{-1}$ in the $H$-band has been measured for the QUIRC 
camera on Gemini by F. Rigaut. These plate scales were combined with 
the position angles of the instruments and the pixel offsets to compute the 
offsets in right ascension and declination that are listed in 
Table~\ref{tab:cand}.

\subsection{Measurement of Detection Limits}

\begin{figure}
\epsscale{0.7}
\plotone{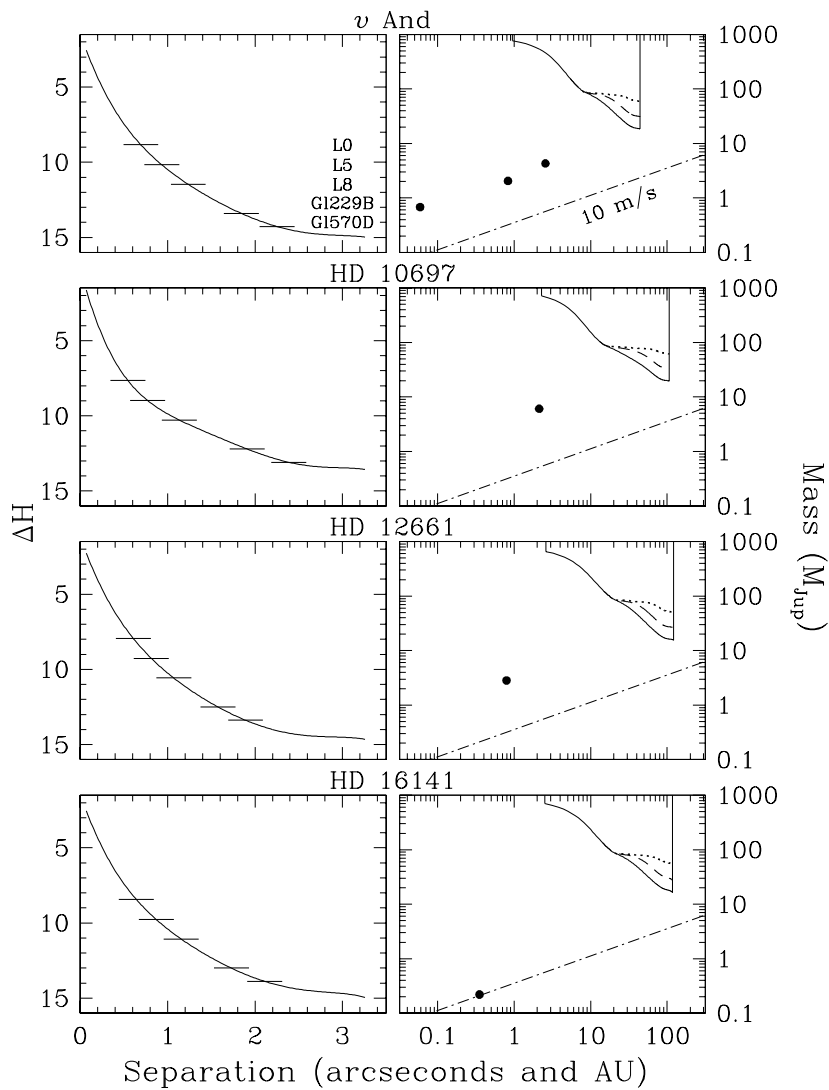}
\caption{
{\it Left:} The solid curves represent the 6~$\sigma$ detection limits for
point sources in Keck AO images of 
$\upsilon$~And, HD~10697, HD~12661, and HD~16141 
as a function of angular separation and after subtracting the $H$-band 
magnitude of each star. 
The short horizontal lines along the detection limits indicate the magnitudes
of typical L0, L5, and L8 field dwarfs and the T dwarfs Gl~229B and Gl~570D 
placed at the distances of these stars (Kirkpatrick et al.\ 2000). 
{\it Right:}
The measured detection limits from the left are plotted in terms of mass
and projected physical separation assuming ages of 1 ({\it solid}), 
3 ({\it dashed}), and 10~Gyr ({\it dotted}) (Burrows et al.\ 1997; 
Baraffe et al.\ 1998). The solid points represent the values of $M$~sin~$i$ 
and the semi-major axis of companions to these stars as measured in radial 
velocity studies. 
Recent stellar age estimates and additional constraints on the companion
masses are found in Table~\ref{tab:prop}. 
A velocity modulation of 10~m~s$^{-1}$ for a solar-mass central star is shown 
as the dot-dashed line. 
}
\label{fig:p1}
\end{figure}

\begin{figure}
\epsscale{0.7}
\plotone{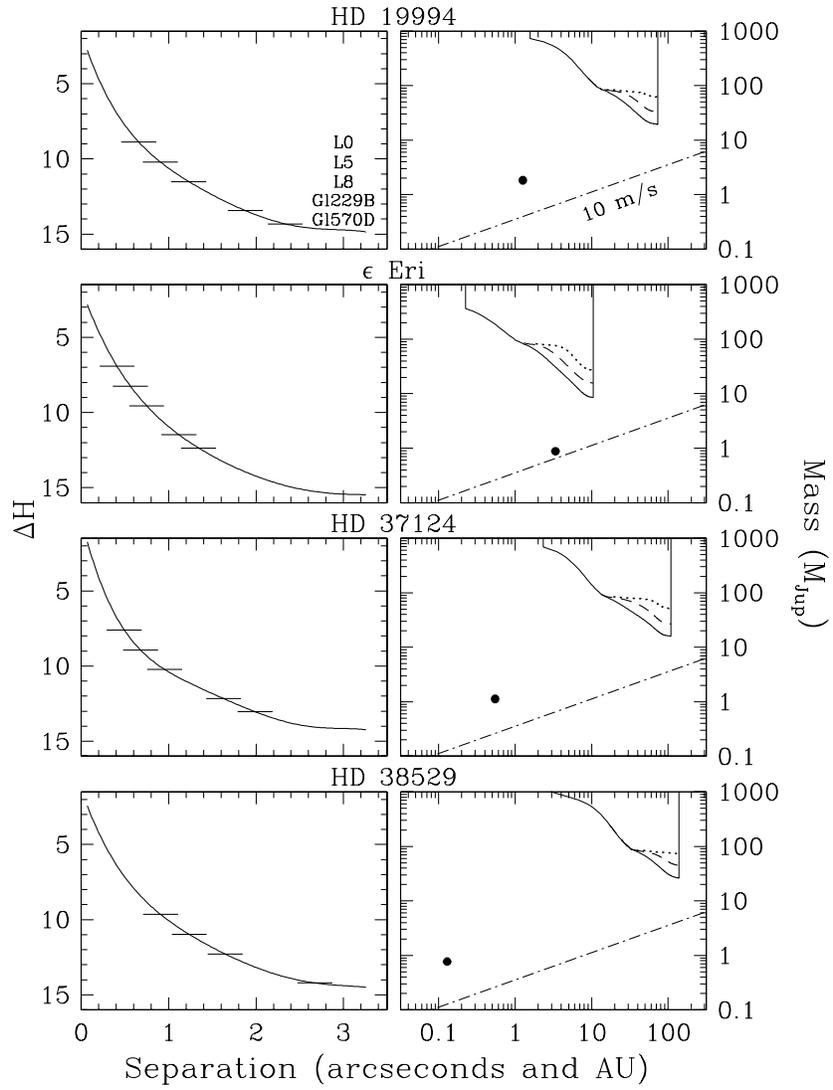}
\caption{
Same as Figure~\ref{fig:p1}, but for 
HD~19994, $\epsilon$~Eri, HD~37124, and HD~38529.
Gl~570D would not be detected in the data for HD~38529.
}
\label{fig:p2}
\end{figure}

\begin{figure}
\epsscale{0.7}
\plotone{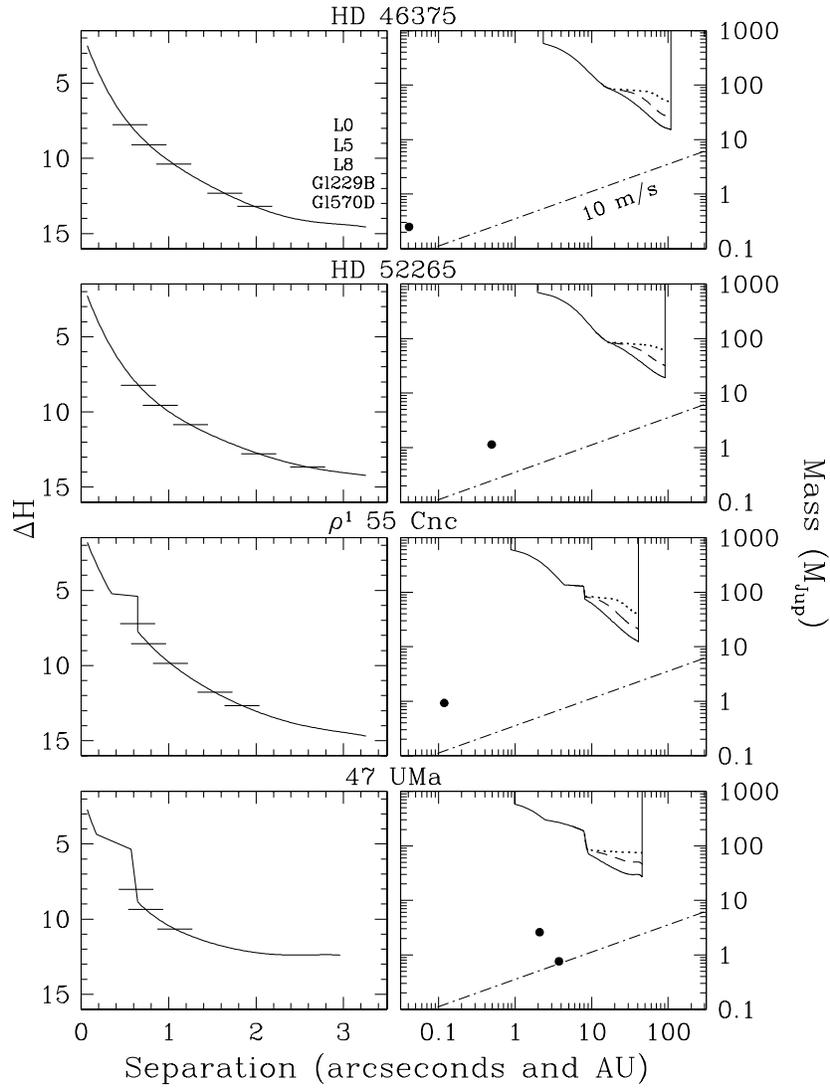}
\caption{
Same as Figure~\ref{fig:p1}, but for 
HD~46375, HD~52265, $\rho^1$~55~Cnc, and 47~UMa.
In the data for 47~UMa, Gl~229B would fall at the detection limit at 
separations beyond $2\arcsec$ and Gl~570D would not be detected.
}
\label{fig:p3}
\end{figure}

\begin{figure}
\epsscale{0.7}
\plotone{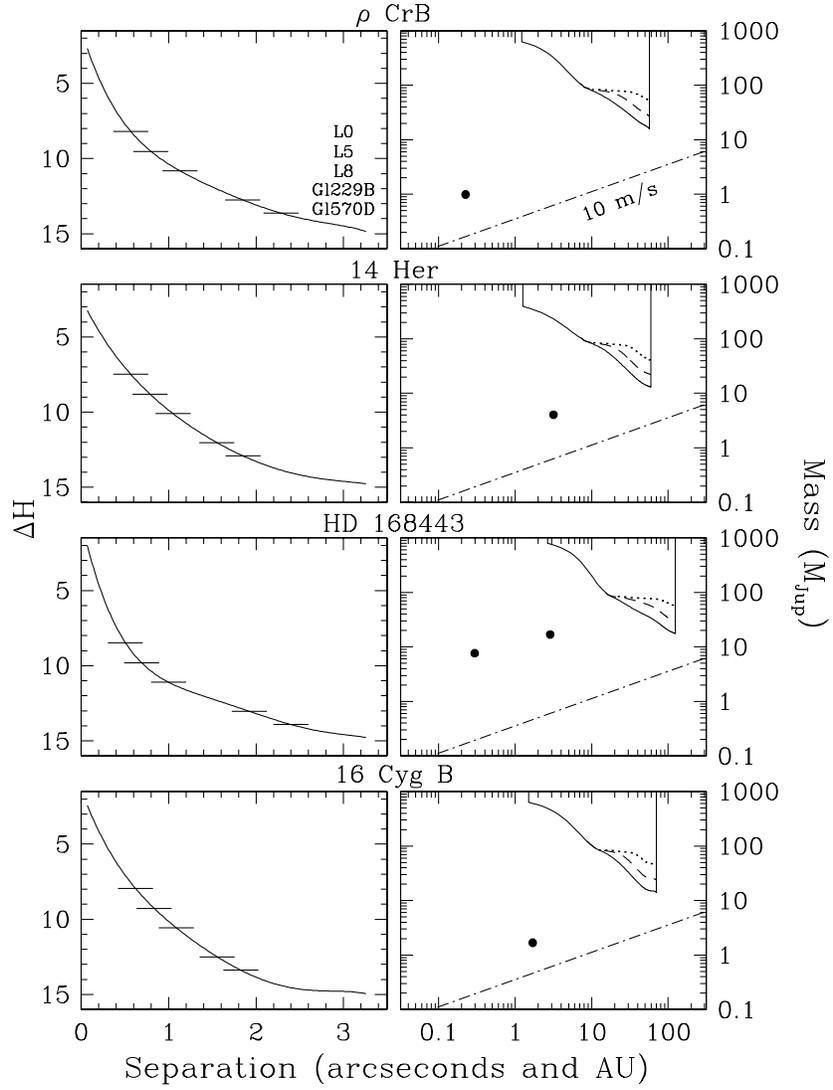}
\caption{
Same as Figure~\ref{fig:p1}, but for
$\rho$~CrB, 14~Her, HD~168443, and 16~Cyg~B.
}
\label{fig:p4}
\end{figure}

\begin{figure}
\epsscale{0.7}
\plotone{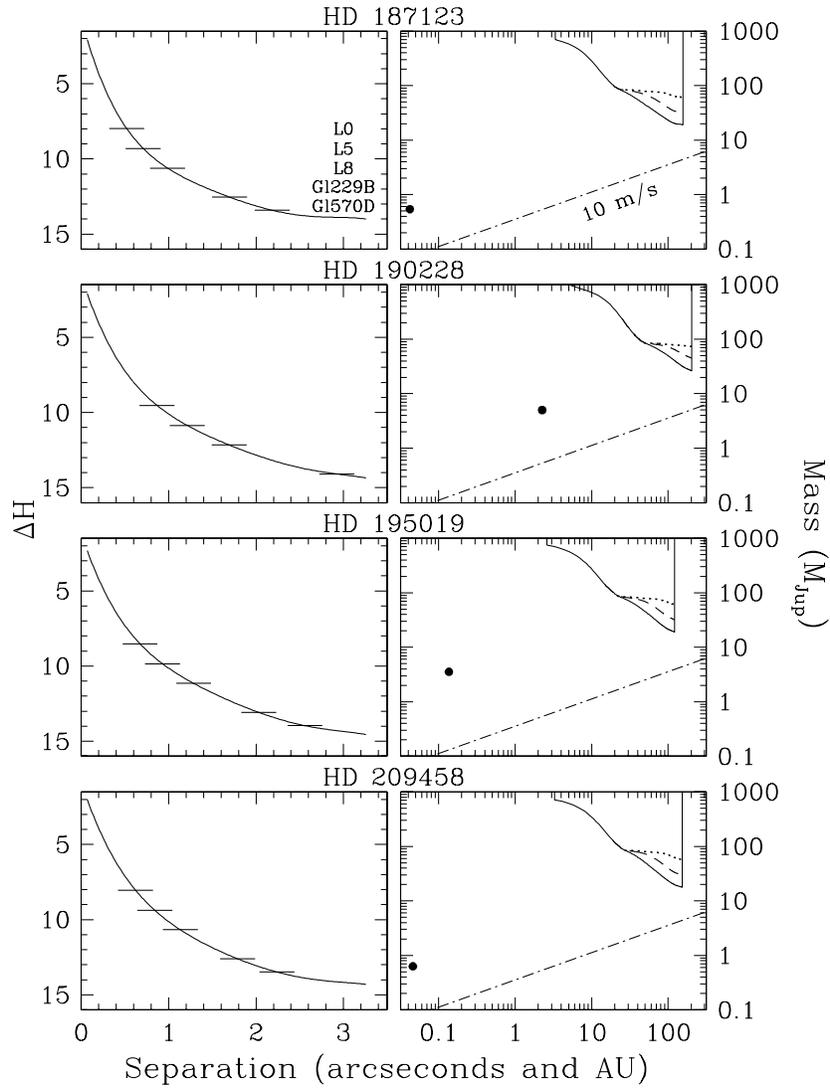}
\caption{
Same as Figure~\ref{fig:p1}, but for
HD~187123, HD~190228, HD~195019, and HD~209458.
Gl~570D would not be detected in the data for HD~190228.
}
\label{fig:p5}
\end{figure}

\begin{figure}
\epsscale{0.7}
\plotone{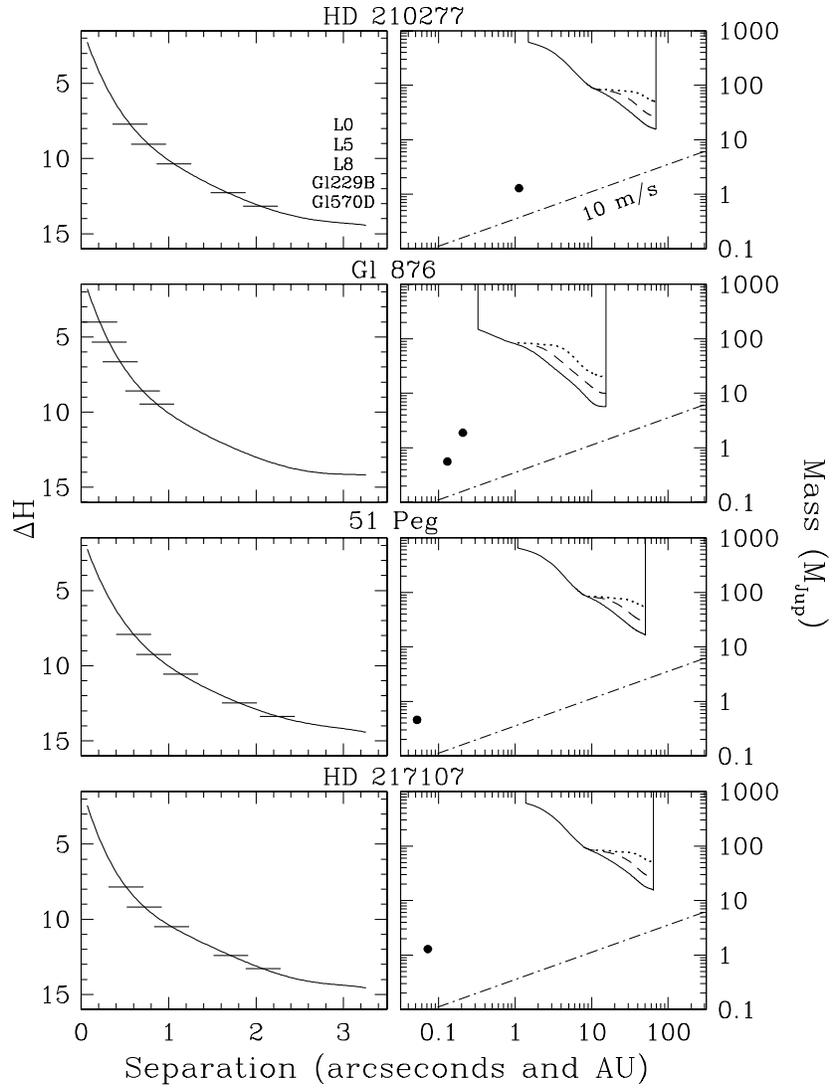}
\caption{
Same as Figure~\ref{fig:p1}, but for
HD~210277, Gl~876, 51~Peg, and HD~217107.
}
\label{fig:p6}
\end{figure}

\begin{figure}
\epsscale{0.7}
\plotone{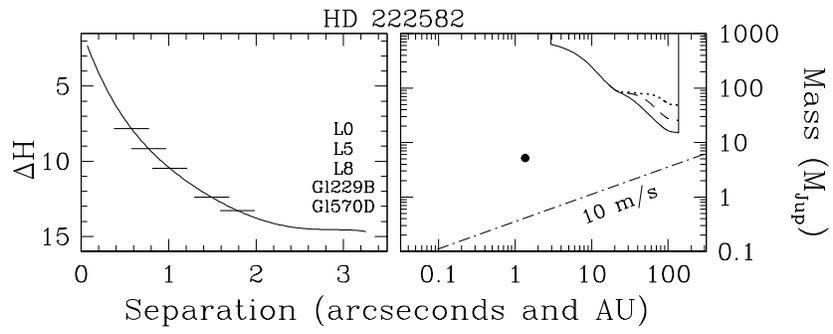}
\caption{
Same as Figure~\ref{fig:p1}, but for
HD~222582.
}
\label{fig:p7}
\end{figure}

We now characterize the sensitivity to companions of the Keck AO data.
We began by measuring the $H$-band detection limits for point sources as a 
function of angular separation from the targets.
To remove the large-scale shape of the PSF of the target stars,
each image was convolved with a two-dimensional Gaussian function and divided 
into the original image. In the quotient image, we 
measured the standard deviation in five pixel square boxes at 
one pixel intervals along a line radially outward from star, which was then
repeated for radial lines at several other angles around the star that avoided
ghost images of the primary.
At the image positions measured in this way, we inserted artificial stars 
that simulated the core of the PSF, which were given the FWHM 
measured from the unsaturated image of the star and a peak intensity
that equaled a multiple of the standard deviation measured at that point 
in the image. After visually examining the artificial stars at several 
values of this multiple, we adopted six sigma as a reasonable
representation of the detection limit. The sources detected around HD~190228
in Figure~\ref{fig:hd190228} are near this detection limit. In addition, eight
artificial stars have been placed at the detection limit of $\Delta H=10$ at
$1\arcsec$ in one of the images of 51~Peg in Figure~\ref{fig:mosaic}.
Within the seeing disc, the presence of speckles complicates the 
characterization of the detection limit. As discussed in the previous section,
real point sources can be identified down to the amplitude of the speckles. As
a result, the speckles effectively define the detection limit for point 
sources. Because the six sigma deviations measured from 5~pixel
boxes in the seeing disc were comparable to the speckles intensities
(i.e., the speckles are sparsely distributed relative to this box size), we
adopted the former measurement for characterizing the detection limit both
inside and outside of the seeing disc. For a given star, the $H$-band
detection limits were measured for the various filters, exposure times, and
position angles and the combined results were fit by one polynomial function. 
The differences of the $H$-band detection limits for point sources and the
magnitudes of the target stars ($\Delta H$) are presented as a function of
angular separation in Figs.~\ref{fig:p1}-\ref{fig:p7}. 
The ghost images that appear in the long exposures result in an 2-3\% 
incompleteness in areal coverage. Because the observations of $\rho^1$~55~Cnc 
and 47~UMa were not optimized for dynamic range, the images were less sensitive
to companions at intermediate separations, as indicated by the
data for these stars in Figure~\ref{fig:p3}. In addition, the
detection limit at large separations for 47~UMa was worse than for the other
stars because only short exposures were obtained.

As an illustration of the sensitivity to companions of these data, 
in Figs.~\ref{fig:p1}-\ref{fig:p7} we have indicated the magnitudes of 
typical L and T dwarfs at the distances of the targets. We considered 
all field dwarfs between spectral types of M9~V and T that have trigonometric 
parallaxes as compiled by Kirkpatrick et al.\ (2000). For these objects, 
we fit a polynomial function to spectral type versus $M_H$ for M9~V through 
L8~V, where the data were from Kirkpatrick et al.\ (2000) and references 
therein. The values of $M_H$ from this function for L0~V, L5~V, and L8~V and 
the measurements of $M_H$ for the T dwarfs Gl~229B and Gl~570D (Kirkpatrick et 
al.\ 2000; Leggett et al.\ 1999) are plotted along the detection limits of 
separation versus $\Delta H$ in Figs.~\ref{fig:p1}-\ref{fig:p7}.

The $H$-band detection limits for companions have been transformed to units of 
mass and projected physical separation in the following manner. 
Conversions between spectral types and effective temperatures are from
Schmidt-Kaler (1982) for $\leq$M0~V and from Leggett et al.\ (1996) as presented
by Luhman (1999) for M1~V to M9~V. For L and T types, we adopted a conversion 
in which L0~V, L8~V, and the T dwarf Gl~229B correspond to effective 
temperatures of 2000, 1400, and 900~K (Leggett et al.\ 1999, 2001; 
Kirkpatrick et al.\ 2000; Reid et al.\ 2001b; Schweitzer et al.\ 2001). 
Bolometric corrections are from Kenyon \& Hartmann (1995) for
$<$M6 and from Bessell (1991), Monet et al.\ (1992), Tinney,
Mould, \& Reid (1993), and Leggett et al.\ (1996) for M6 to M9.
For L and T types, the $H$-band bolometric corrections were computed from
the $J$-band corrections of Reid et al.\ (2001a) and the average $J-H$ colors 
of Kirkpatrick et al.\ (2000).
By combining these conversions with the $H$-band magnitudes and angular 
separations of the detection limits, the distances of the targets, and
theoretical mass-luminosity relations for 1, 3, and 10~Gyr 
($\geq0.1$~\msun, Baraffe at al.\ 1998; $<0.1$~\msun, Burrows et al.\ 1997), 
we have expressed the detection limits for companions in terms of mass and
projected physical separation in Figs.~\ref{fig:p1}-\ref{fig:p7}.
For comparison, we also plot the positions of the companions to these
stars that have been discovered through radial velocity measurements. 
A typical detection limit of those studies is represented by a velocity 
modulation of 10~m~s$^{-1}$ for 1~\msun. The actual detection limits for many
of these stars are presented by Cumming, Marcy, \& Butler (1999).

\section{Discussion}

We have obtained Keck near-IR AO images of 25 of 
the $\sim60$ known extrasolar planetary systems. From these data, we have
identified faint sources near four of the targets and have measured the
detection limits for companions. Using the Keck and Gemini AO systems, 
we have also obtained AO images at an additional epoch of three of the four
primaries exhibiting nearby faint objects. 
We now discuss the candidate companions
individually, evaluate the achieved detection limits, and examine
implications of these new constraints on substellar companions.
For reference during this discussion, we list the 25 planetary systems
in Table~\ref{tab:prop} along with the latest values of separation and
$M$~sin~$i$ for their radial velocity companions, upper limits
to the masses of these companions from {\it Hipparcos} astrometry 
(Zucker \& Mazeh 2001; see also Pourbaix \& Arenou 2001), and recent age 
estimates. 

Faint point sources have been detected in the Keck AO images of HD~37124, 
HD~168443, HD~187123, and HD~190228, which could be either companions or 
background field stars. 
These four star are all located at low galactic latitude ($|b|<6\arcdeg$), 
which suggests that most of the faint sources are background field 
stars rather than companions. 
For the star HD~168443 in particular, which is only $20\arcdeg$ from the 
Galactic center and has $b=2\fdg5$, the presence of nine nearby objects at 
comparable magnitudes ($H\sim17$-19) is suggestive of a background star 
population. 
Because of the high proper motions of the targets in our sample and 
the high spatial resolution of AO imaging, multi-epoch observations spanning 
a year or less can easily distinguish companions from background stars. 
Such data were obtained for five of the 15 candidate companions, as listed
in Table~\ref{tab:cand}. All of these five objects exhibit proper motions
relative to the primary stars that are consistent with those expected of
background stars. 
Between 2000 September 18 and 2001 October 7, the position of the object
near HD~37124 changed by 
($\Delta \alpha$, $\Delta \delta$)=($0\farcs104$, $0\farcs388$)$\pm0\farcs03$,
which is close to the relative motion of ($0\farcs0839\pm0\farcs0014$, 
$0\farcs4418\pm0\farcs0009$) expected for a stationary background star given 
the {\it Hipparcos} proper motion of HD~37124.
The two objects near HD~187123 were observed at two epochs by Keck. In
data obtained at a third epoch by Gemini, the fainter source fell below the
detection limit. 
The position of the brighter object changed by 
($-0\farcs193$, $0\farcs157$)$\pm0\farcs03$ relative to HD~187123 from 2000 
June 27 to 2001 October 7, which is consistent with the motion of
($-0\farcs1831\pm0\farcs0007$, $0\farcs1577\pm0\farcs0008$) expected for 
a stationary background star.
From 2000 June 27 to September 17, the offsets between the two candidate
companions to HD~187123 are expected to have changed by $\sim0\farcs03$
in right ascension and declination if one was a background star and the
other was a companion. Since the sources remained fixed relative to each other
at a level of $\lesssim0\farcs005$, the fainter source also must be background 
star.  Multi-epoch data were obtained for two of the sources near HD~168443. 
Between 2000 May 10 and September 17, the positions 
of two of the objects relative to HD~168443 changed by 
($0\farcs078$, $0\farcs078$)$\pm0\farcs03$
and ($0\farcs077$, $0\farcs085$)$\pm0\farcs03$. These measurements are
consistent with the motion of background stars, which is expected 
to be ($\Delta \alpha$, $\Delta \delta$)=($0\farcs033$, $0\farcs080$) given 
the proper motion of HD~168443. 
Because of the low galactic latitude of HD~168443 and HD~190228 
($|b|<3\arcdeg$), the remaining faint sources detected near these two stars 
are probably background stars as well. 

Rather than simply the discovery of low-mass companions, the primary objective
of this study has been to obtain new, highly sensitive constraints on
the presence of companions near a sample of planetary systems. These 
constraints are summarized and quantified by the point source detection 
limits in Figs.~\ref{fig:p1}-\ref{fig:p7}. From the plots of separation versus 
$\Delta H$, we see that AO on the Keck~II telescope produces
high-contrast images in which point sources at $\Delta H=10$ can be 
detected at $1\arcsec$ from a bright star (see Figure~\ref{fig:mosaic}).
As a result, these data are 
sensitive to L and T dwarf companions to the targets at separations from 
$0\farcs5$ (L0) to $2\arcsec$ (Gl~570D). For comparison, the companion search
towards field stars by Oppenheimer et al.\ (2001) probed 3-$30\arcsec$ and 
discovered the brown dwarf Gl~228B at a separation of $7\farcs8$.
In recent coronographic AO images of the planet-bearing star Gl~86, 
Els et al.\ (2001) identified a probable brown dwarf companion at $1\farcs72$, 
$H=14.4$, and $\Delta H=10$. Gl~86 and $\rho^1$~55~Cnc have similar distances 
and brightnesses, and our detection limit for the latter indicates that the 
companion to Gl~86 would have been detected in our program as well.
When these detection limits are transformed to units of companion mass and
physical separation, they can be compared to the sensitivities of other 
methods of searching for substellar companions. The most fruitful of these
techniques has been radial velocity monitoring, which currently reaches
companions at $\lesssim5$~AU that induce modulations of $\gtrsim10$~m~s$^{-1}$.
As demonstrated in Figs.~\ref{fig:p1}-\ref{fig:p7}, our AO survey nicely 
complements the radial velocity work by probing for companions at wider 
separations, typically stars at 3-10~AU and brown dwarfs at 10-100~AU.
The sensitivity to companions in direct imaging is improved with less 
massive and closer primary stars, which is illustrated by our data for
the M dwarf Gl~876, where brown dwarfs are detectable at 1-10~AU.

Several stars in our sample deserve further comment given the 
new constraints on the presence of companions from this study.
As already noted, the AO detection limits approach but do not reach the 
separations and masses of the substellar objects that have been inferred from
radial velocity measurements to date. However, in a few cases, the AO data 
do place upper limits on the masses of those unseen companions. 
On the dates of our observations, radial velocity measurements predict 
minimum angular separations of $0\farcs04$ and $0\farcs231$ for the two 
companions to 47~UMa and $0\farcs148$ for the companion to 14~Her (G. Marcy, 
private communication). 
While we cannot constrain the mass of the inner companion to 47~UMa, 
our non-detections correspond to a mass upper limit of 0.3~\msun\ for each
of the other two companions.
The currently measured orbit for the putative companion to $\epsilon$~Eri is too
uncertain to reliably predict the minimum angular separation, but our data do 
suggest that the companion is unlikely to be a star.
As the time baselines of the radial velocity studies become longer, additional
companions may be discovered at larger separations (see Table~\ref{tab:prop}),
and these AO data could be useful in constraining their masses as well.
However, upper limits from {\it Hipparcos} are often much lower.
Finally, Gonzales et al.\ (2001) found that HD~46375 and HD~37124
are overluminous on the Hertzsprung-Russell diagram by 0.5 and 0.25~mag at 
$V$, respectively. They suggest that these stars may have unresolved
stellar companions that are beyond the maximum separations of 5~AU
probed by radial velocity measurements. However, at projected separations
of $>0\farcs1$ ($>3.3$~AU for each star), we find no companions that are
sufficiently bright to account for the discrepancies in the luminosities.
Other constraints on the presence of companions from the literature are
noted in Table~\ref{tab:prop}.

\section{Conclusion}

We have conducted a survey for companions to 25 of the $\sim60$ known
extrasolar planetary systems using near-IR AO imaging at Keck Observatory.
The Keck~II AO system has produced high-resolution images that approach
the diffraction limit with a typical value of $0\farcs04$ for the FWHM
in the $H$ band. Using a variety of neutral density filters, 
exposure times, and position angles, we arrived at a set of images for
each target that provides optimum sensitivity for all separations from 
$0\farcs1$ to $3\farcs3$. 

After searching the AO images for candidate companions, 
near HD~37124, HD~168443, HD~187123, and HD~190228 we have identified 
15 faint point sources ($H=14$-20), all of which would be at or below
the hydrogen burning mass limit if they were companions. 
Using AO imaging at Keck and Gemini Observatories, we have obtained images
at a second epoch for five of the brightest candidate companions. These
sources all exhibit proper motions relative to the central stars that
are consistent with those expected of background stars. 
The remaining faint sources are probably background stars as well given the low 
galactic latitude of the four targets.

We have measured the point source detection limits for the AO images, which
are presented in terms of both angular separation versus $\Delta H$
and projected physical separation versus companion mass.
The sensitivities correspond to $\Delta H=10$ at $1\arcsec$ and L and T 
dwarfs at $0\farcs5$ to $2\arcsec$ from the planet-bearing stars ($H=2$-6).
These data reach companion stars and brown dwarfs at 3-10~AU and 10-100~AU
from the planet-bearing stars in this study, providing an essential complement 
to previous radial velocity measurements of these systems ($>0.3$~\mj, $<5$~AU).

\acknowledgements
We wish to thank Peter Wizinowich, Scott Acton, David Le Mignant and the rest
of the Keck Observatory staff for their support. We also thank Kathy
Roth, Francois Rigaut, Mark Chun, Olivier Guyon, and Dan Potter for their 
assistance in obtaining the observations at Gemini Observatory. We are 
grateful to Joan Najita for help with an earlier Keck proposal, and to Geoff 
Marcy for calculating the positions of the radial velocity companions to 47 UMa
and 14 Her. We thank Geoff Marcy, Debra Fischer, and Jamie Lloyd for useful 
discussions. K. L. was supported by a postdoctoral fellowship at the 
Harvard-Smithsonian Center for Astrophysics. R. J. was supported by a Miller 
Research Fellowship. 
This work was supported in part by the Smithsonian Institution, NASA, and NSF 
grants to Geoff Marcy, and a NASA grant to R.J. administered by the AAS.
This publication makes use of data products from the Two Micron All Sky Survey,
which is a joint project of the University of Massachusetts and the
Infrared Processing and Analysis Center, funded by the National Aeronautics and
Space Administration and the National Science Foundation.
Some of the data presented herein were obtained at
the W. M. Keck Observatory, which is operated as a scientific partnership
among the California Institute of Technology, the University of California,
and the National Aeronautics and Space Administration.
The Observatory was made possible by the generous financial support of
the W. M. Keck Foundation.
We wish to extend special thanks to those of Hawaiian ancestry on whose sacred
mountain we are privileged to be guests. Without their generous hospitality,
some of the observations presented herein would not have been possible.
Some of these observations were obtained at the Gemini Observatory, 
which is operated by the Association of Universities for Research in Astronomy, 
Inc., under a cooperative agreement with the NSF on behalf of the Gemini 
partnership: the National Science Foundation (United States), the Particle 
Physics and Astronomy Research Council (United Kingdom), the National Research 
Council (Canada), CONICYT (Chile), the Australian Research Council (Australia), 
CNPq (Brazil) and CONICET (Argentina).
This paper includes data obtained with the Adaptive Optics System 
Hokupa'a/Quirc, developed and operated by the University of Hawaii Adaptive 
Optics Group, with support from the National Science Foundation.
Finally, we thank the anonymous referee for useful comments that improved this
manuscript.

\newpage

\clearpage
 
\clearpage
 
\clearpage
 
\clearpage
 
\clearpage
 
\clearpage
 
\clearpage
 
\clearpage
 
\clearpage
 
\clearpage
 
\clearpage
 
\clearpage
 
\clearpage
 
\begin{deluxetable}{lllllll}
\tabletypesize{\scriptsize}
\tablewidth{0pt}
\tablecaption{Properties of Planetary Systems \label{tab:prop}}
\tablehead{
\colhead{Star} &
\colhead{$a$\tablenotemark{a}} &
\colhead{$M$~sin~$i$\tablenotemark{a}} &
\colhead{References} &
\colhead{$M_{Hip}$\tablenotemark{b}} &
\colhead{Stellar Age} &
\colhead{References} \\
\colhead{} &
\colhead{(AU)} &
\colhead{(\mj)} &
\colhead{} &
\colhead{(\mj)} &
\colhead{(Gyr)} &
\colhead{} 
}
\startdata
$\upsilon$~And & 0.059 & 0.58 & 1 & \nodata & $3.8\pm1$, $2.7\pm0.3$, $2.6^{+2.1}_{-1}$, $2.9\pm0.6$, $3.3\pm0.5$, $<4$ & 2, 3, 4, 5, 6, 7 \\
 & 0.828 & 2.05 & 8, 9 & \nodata & & \\
 & 2.56  & 4.29 & 8, 9 & $<19$ & & \\
HD~10697 & 2.12 & 6.08 & 10 & $<92$ & $7.8\pm0.5$, 6.0 & 11 \\
HD~12661 & 0.795 & 2.84 & 12 & $<140$ & $8\pm1$, 8.4 & 11 \\
HD~16141 & 0.351 & 0.22 & 13 & $<530$ & $8.5\pm0.5$, 6.7 & 11 \\
HD~19994 & 1.26 & 1.83 & 14 & $<66$ & $3\pm0.3$, 4.3 & 3, 15 \\
$\epsilon$~Eri\tablenotemark{c} & 3.36 & 0.88 & 16 & $<22$ & 0.5-1, $0.73\pm0.2$ & 17, 18 \\
HD~37124 & 0.547 & 1.13 & 10 & $<480$ & 3.9 & 11 \\
HD~38529\tablenotemark{d} & 0.129 & 0.77 & 9 & \nodata & $3\pm0.5$, 3.7 & 11 \\
HD~46375 & 0.041 & 0.25 & 13 & \nodata & 4.5 & 11 \\
HD~52265 & 0.493 & 1.14 & 12, 19 & $<140$ & $2.1\pm0.3$, 4.0, 3.5 & 8, 8, 19 \\
$\rho^1$~55~Cnc\tablenotemark{e} & 0.118 & 0.93 & 1 & $<290$ & $<5$, $4\pm3$, 5, $8.4^{+7.1}_{-8.3}$, 4-5 & 2, 20, 21, 4, 7 \\
47~UMa & 2.09 & 2.54 & 22 & $<15$ & $7.3\pm1.9$, $6.6^{+1.5}_{-1.3}$, $8\pm2$, 6.3, $6.3^{+2}_{-2.4}$, 6 & 23, 3, 24, 21, 5, 7 \\
 & 3.73 & 0.76 & 25 & \nodata & & \\
$\rho$~CrB & 0.224 & 0.99 & 26 & $<230$ & $10.2\pm1.7$, $12.1\pm1.1$, $11\pm2$, 12.3, $14.1^{+2}_{-2.4}$, 6 & 2, 3, 24, 21, 4, 7 \\
14~Her & 3.17 & 4.05 & 27 & $<20$ & 6 & 28, 21 \\
HD~168443 & 0.295 & 7.64 & 29 & \nodata & $10.5\pm1.5$, 7.4, 8 & 11, 11, 30 \\
 & 2.87 & 16.96 & 30, 31 & $<80$ & & \\
16~Cyg~B & 1.69 & 1.68 & 32 & $<27$ & $8\pm1.8$, $9\pm2$, 9, 6-7 & 2, 24, 21, 7 \\
HD~187123\tablenotemark{f} & 0.042 & 0.54 & 33 & \nodata & $4^{+1}_{-0.5}$, 5.5 & 28, 21 \\
HD~190228 & 2.25 & 5.01 & 34 & $<77$ & \nodata & \nodata \\
HD~195019 & 0.136 & 3.55 & 35 & $<1470$ & 3 & 35 \\
HD~209458 & 0.046 & 0.63\tablenotemark{g} & 36, 37 & \nodata & 4.5, $3\pm1$, 4.3 & 38, 11, 11 \\
HD~210277\tablenotemark{h} & 1.12 & 1.29 & 29, 19 & $<40$ & $12\pm2$, 8.5 & 28, 21 \\
Gl~876 & 0.207 & 1.89 & 39, 40 & $<51$ & 1-10 & 39 \\
 & 0.130 & 0.56 & 41 & \nodata & & \\
51~Peg & 0.052 & 0.46 & 42 & $<1080$ & $4\pm2.5$, $7.0^{+1.1}_{-0.9}$, $7.6^{+4}_{-5.1}$, $5.1^{+3}_{-0.7}$, 7, $5.5\pm0.5$, 7.1 & 23, 3, 4, 5, 7, 11, 11 \\
HD~217107\tablenotemark{i} & 0.072 & 1.29 & 35 & $<1670$ & 8, $12\pm1.5$, 5.6 & 35, 11, 11 \\
HD~222582 & 1.35 & 5.18 & 10 & $<150$ & $11\pm1$, 5.6 & 11 \\
\enddata
\tablenotetext{a}{http://exoplanets.org.}
\tablenotetext{b}{Constraints on companion masses derived from 
{\it Hipparcos} data and radial velocity measurements (Zucker \& Mazeh 2001).
}
\tablenotetext{c}{Companion has a probable mass of $\sim1.7$~\mj\ (Hatzes et 
al.\ 2000).}
\tablenotetext{d}{May have an additional companion (Fischer et al.\ 2001b).}
\tablenotetext{e}{May have an additional companion at $\sim5.5$~AU and 
$\sim3.2$~\mj\ (Butler et al.\ 1997; Fischer et al.\ 2001b).}
\tablenotetext{f}{May have an additional companion (Vogt et al.\ 2000).}
\tablenotetext{g}{$i=86\fdg6\pm0\fdg14$ (Brown et al.\ 2001).}
\tablenotetext{h}{No stellar companions at $<10$~AU and 17-250~AU
(Marcy et al.\ 1999, references therein).}
\tablenotetext{i}{Companion has a probable mass upper limit of
$\sim11$~\mj\ (Fischer et al.\ 1999). May have an additional companion 
beyond a few AU and more massive than $\sim3$~\mj\ (Vogt et al.\ 2000; 
Fischer et al.\ 2001b).}
\tablerefs{
(1) Butler et al.\ 1997; 
(2) Fuhrmann et al.\ 1998;
(3) Ng \& Bertelli 1998;
(4) Ford et al.\ 1999;
(5) Lachaume et al.\ 1999;
(6) Gonzales \& Laws 2000;
(7) Henry et al.\ 2000a;
(8) Butler et al.\ 1999;
(9) Fischer et al.\ 2001b;
(10) Vogt et al.\ 2000;
(11) Gonzales et al.\ 2001;
(12) Butler et al.\ 2000;
(13) Marcy, Butler, \& Vogt 2000a;
(14) Queloz et al.\ 2001;
(15) Smith, Cunha, \& Lazzaro 2001;
(16) Hatzes et al.\ 2000;
(17) Greaves et al.\ 1998;
(18) Song et al.\ 2000;
(19) Naef et al.\ 2001;
(20) Gonzalez \& Vanture 1998;
(21) Gonzalez 1999;
(22) Marcy \& Butler 1996;
(23) Fuhrmann, Pfeiffer, \& Bernkopf 1997;
(24) Gonzalez 1998;
(25) Fischer et al.\ 2001a;
(26) Noyes et al.\ 1997;
(27) Marcy et al.\ 2000b;
(28) Gonzalez et al.\ 1999;
(29) Marcy et al.\ 1999;
(30) Marcy et al.\ 2001b;
(31) Udry, Mayor, \& Queloz 2001;
(32) Cochran et al.\ 1997;
(33) Butler et al.\ 1998;
(34) Sivan et al.\ 2001;
(35) Fischer et al.\ 1999;
(36) Charbonneau et al.\ 2000;
(37) Henry et al.\ 2000b;
(38) Mazeh et al.\ 2000;
(39) Marcy et al.\ 1998;
(40) Delfosse et al.\ 1998;
(41) Marcy et al.\ 2001a;
(42) Mayor \& Queloz 1995.
}
\end{deluxetable}

\clearpage

\begin{deluxetable}{llllll}
\tablewidth{0pt}
\tablecaption{Faint Sources Detected Near Planetary Systems \label{tab:cand}}
\tablehead{
\colhead{Star} &
\colhead{$\Delta\alpha$\tablenotemark{a}} &
\colhead{$\Delta\delta$\tablenotemark{a}} &
\colhead{$H$} &
\colhead{$\Delta H$\tablenotemark{b}} &
\colhead{Obs Date} \\
\colhead{} & \colhead{(arcsec)} & \colhead{(arcsec)} & \colhead{(mag)}
& \colhead{(mag)} & \colhead{}
}
\startdata
HD~37124 & $-$3.076, $-2.972$ & 0.314, 0.702 & 14.9 & 8.9 & 9/18/00, 10/07/01 \\
HD~168443 & 2.802 & $-$0.413 & 18.5 & 13.1 & 9/17/00 \\ 
 & 3.516, 3.594 & 0.129, 0.207 & 16.4 & 11.0 & 5/10/00, 9/17/00 \\
 & 3.166, 3.243 & 0.318, 0.403 & 17.7 & 12.3 & 5/10/00, 9/17/00 \\
 & 2.691 & 1.585 &  18.8 & 13.4 & 9/17/00 \\
 & 1.554 & 2.789 &  19.0 & 13.6 & 9/17/00 \\
 & 0.402 & 2.677 &  17.6 & 12.2 & 9/17/00 \\
 & $-$0.364 & 2.949 &  18.6 & 13.2 & 9/17/00 \\
 & $-$3.215 & 2.380 & 19.0 & 13.6 & 9/17/00 \\
 & 2.973 & 1.502  &  $\sim20$ & $\sim14.6$ & 9/17/00 \\
HD~187123 & 3.364,3.361,3.171 & 0.973,0.994,1.130 & 14.2 & 7.8 & 6/27/00, 9/16/00, 10/07/01 \\
 & 3.142,3.137 & 0.286,0.310 & 18.3 & 11.9 & 6/27/00, 9/16/00 \\
HD~190228 & 0.999 & $-$1.796 & $\sim19$ & $\sim13.6$ & 9/18/00 \\
 & $-$1.343 & $-$1.644 & $\sim18.5$ & $\sim13.1$ & 9/18/00 \\
 & $-$1.664 & $-$2.312 & $\sim19$ & $\sim13.6$ & 9/18/00 \\
\enddata
\tablenotetext{a}{Positions in right ascension and declination are measured
with respect to the star. For a given star, the uncertainties in the positions 
of the faint sources relative to the star and relative to each other are 
$\pm0\farcs02$ and $\pm0\farcs004$, respectively. An exception is HD~190228, 
where the uncertainty in the former value is $\pm0\farcs04$.}
\tablenotetext{b}{Magnitude difference of the primary star and faint source.}
\end{deluxetable}

\end{document}